\documentclass[prd,aps,nofootinbib,preprintnumbers]{revtex4}
\usepackage{amsmath,amssymb,epsf}
\usepackage{graphicx, pstricks}
\begin{document}
\newcommand{\lefile}[1]{{\ttfamily #1}}
\newcommand{\levariable}[1]{{\ttfamily \slshape #1}}
\newcommand{\lefunction}[1]{{\ttfamily \slshape #1()}}
\input epsf
\renewcommand{\topfraction}{0.8}
\title { \Large  \bf  CLUSTEREASY\\A Program for Simulating Scalar
  Field Evolution on Parallel Computers}
\author{Gary Felder}
\affiliation{Department of Physics, Smith College, Northampton, MA
  01063, USA}
\date{\today}
\begin{abstract}
We describe a new, parallel programming version of the scalar field
simulation program LATTICEEASY. The new C++ program, CLUSTEREASY, can
simulate arbitrary scalar field models on distributed-memory
clusters. The speed and memory requirements scale well with the number
of processors. As with the serial version of LATTICEEASY, CLUSTEREASY
can run simulations in one, two, or three dimensions, with or without
expansion of the universe, with customizable parameters and
output. The program and its full documentation are available on the
LATTICEEASY website at
http://www.science.smith.edu/departments/Physics/fstaff/gfelder/latticeeasy/. In
this paper we provide a brief overview of what CLUSTEREASY does and
the ways in which it does and doesn't differ from the serial version
of LATTICEEASY.
\end{abstract}
\maketitle

\section{Introduction}

Studying the early universe requires describing the evolution of
interacting fields in a dense, high-energy environment. The study of
reheating after inflation and the subsequent thermalization of the
fields produced in this process typically involves non-perturbative
interactions of fields with exponentially large occupations numbers in
states far from thermal equilibrium. Various approximation methods
have been applied to these calculations, including linearized analysis
and the Hartree approximation. These methods fail, however, as soon as
the field fluctuations become large enough that they can no longer be
considered small perturbations. In such a situation linear analysis no
longer makes sense and the Hartree approximation neglects important
rescattering terms. In many models of inflation preheating can amplify
fluctuations to these large scales within a few oscillations of the
inflaton field. Moreover, such large amplification appears to be a
generic feature, arising via parametric resonance in single-field
inflationary models and tachyonic instabilities in hybrid models.

The only way to fully treat the nonlinear dynamics of these systems is
through lattice simulations. These simulations directly solve the
classical equations of motion for the fields. Although this approach
involves the approximation of neglecting quantum effects, these
effects are exponentially small once preheating begins. So in any
inflationary model in which preheating can occur lattice simulations
provide the most accurate means of studying post-inflationary
dynamics.

In 2000 G.F.and Igor Tkachev released LATTICEEASY \cite{latticeeasy},
a C++ program for simulating scalar field evolution in an expanding
universe. In the ensuing years LATTICEEASY has been used by us and
other groups to study such topics as preheating, baryogenesis, gravity
wave production, and more. These simulations have been extremely
useful, but they have for the most part been confined to relatively
simple toy models, primarily due to computational limitations. To
study cosmology in more complex models such as the MSSM or GUT
theories will require the use of large, parallel clusters. CLUSTEREASY
is a version of LATTICEEASY that can be run in parallel on multiple
processors.

Section \ref{overview} of this paper gives a brief overview of what
LATTICEEASY does and how to use it, and notes the modifications that
must be made in the LATTICEEASY files to run them in
CLUSTEREASY. Section \ref{implementation} describes the algorithms
used to parallelize the simulations. For more detailed documentation
the reader is referred to the LATTICEEASY website
\newline
\verb|http://www.science.smith.edu/departments/Physics/fstaff/gfelder/latticeeasy/|
\newline

\section{Overview}\label{overview}

LATTICEEASY consists of several C++ files, but only two are designed
to be modified by most users. Each particular scalar field potential
that the program solves is encoded in a {\it model file} called
\lefile{model.h}, in which the user enters equations for the potential
and its various derivatives. The parameters that control individual
runs are stored in a file called \lefile{parameters.h}. These
parameters include physical quantities such as masses and couplings,
numerical quantities such as the number of gridpoints and the size of
the time step, and parameters to control what types of output are
generated by the simulation.

To use CLUSTEREASY the user must replace all of the
non-user-modifiable files from LATTICEEASY with the new, CLUSTEREASY
versions. The parameter files from LATTICEEASY can be used with no
changes, however, and the model files only need the addition of two
lines, described in the online documentation.

The formats of the outputs created by CLUSTEREASY are the same as
those from the serial version. One of the output options is to create
a grid image that can be used to resume a run and continue it to later
times. Grid images created by LATTICEEASY can be read in by
CLUSTEREASY and vice-versa. The only way to distinguish CLUSTEREASY
output from LATTICEEASY output is the file \lefile{info}, which
contains basic information about the run such as the potential used
and the physical and numerical parameters. In CLUSTEREASY this file
has an additional line specifying the number of processors used for
the run.

To run CLUSTEREASY you need a cluster with MPI. MPI is a standard set
of libraries used for parallel programming in C, C++, and Fortran, and
should be installed on any standard cluster. You also need the freely
available Fourier Transform library FFTW. (See the online
documentation for possible compatibility issues with the way FFTW is
installed on different systems and how to resolve them.)

The makefile that comes with CLUSTEREASY assumes that the command for
compiling a C++ MPI program is mpiCC. If this command is different on
your system you will need to modify the makefile
accordingly. Otherwise you should be able to compile the code simply
by typing ``make.'' You should consult your system documentation for
the correct syntax for running a parallel program, but on most
clusters it is
\newline
\verb|mpirun -np <number of processors> latticeeasy|
\newline
Note that the number of processors is determined at execution-time,
not at compile-time.

\section{Parallel Algorithms}\label{implementation}

LATTICEEASY uses a staggered leapfrog algorithm with a fixed time
step. This means that at each step the field values $f$ and their
derivatives $\dot{f}$ are stored at two different times $t$ and
$t+dt/2$ respectively. The derivatives are used to advance the field
values by a full step $dt$ and then the field values are used to
calculate the second derivatives $\ddot{f}$, which are in turn used to
advance the field derivatives by $dt$. This evolution is done in
place, meaning the newly calculated field values and/or derivatives
overwrite the old ones.

To implement this scheme on multiple processors CLUSTEREASY uses
``slab decomposition,'' meaning the grid is divided along a single
dimension (the first spatial dimension). For example, in a 2D run with
$N=8$ on two processors, each processor would cast a $4 \times 8$ grid
for each field. At each processor the variable \levariable{n} stores
the local size of the grid in the first dimension, so in this example
each processor would store $n=4$, $N=8$. Note that \levariable{n} is
not always the same for all processors, but it generally will be if
the number of processors is a factor of \levariable{N}.

In practice, the grids are actually slightly larger than $n \times N$ because
calculating spatial derivatives at a gridpoint requires knowing the
neighboring values, so each processor actually has two additional
columns for storing the values needed for these gradients. Continuing
the example from the previous paragraph, each processor would store a
$6 \times 8$ grid for each field. Within this grid the values $i=0$ and
$i=5$ would be used for storing ``buffer'' values, and the actual
evolution would be calculated in the range $1 \le i \le 4$, $0 \le j
\le 7$.

\begin{figure}[ht]
\centering
\includegraphics[width=0.9\textwidth]{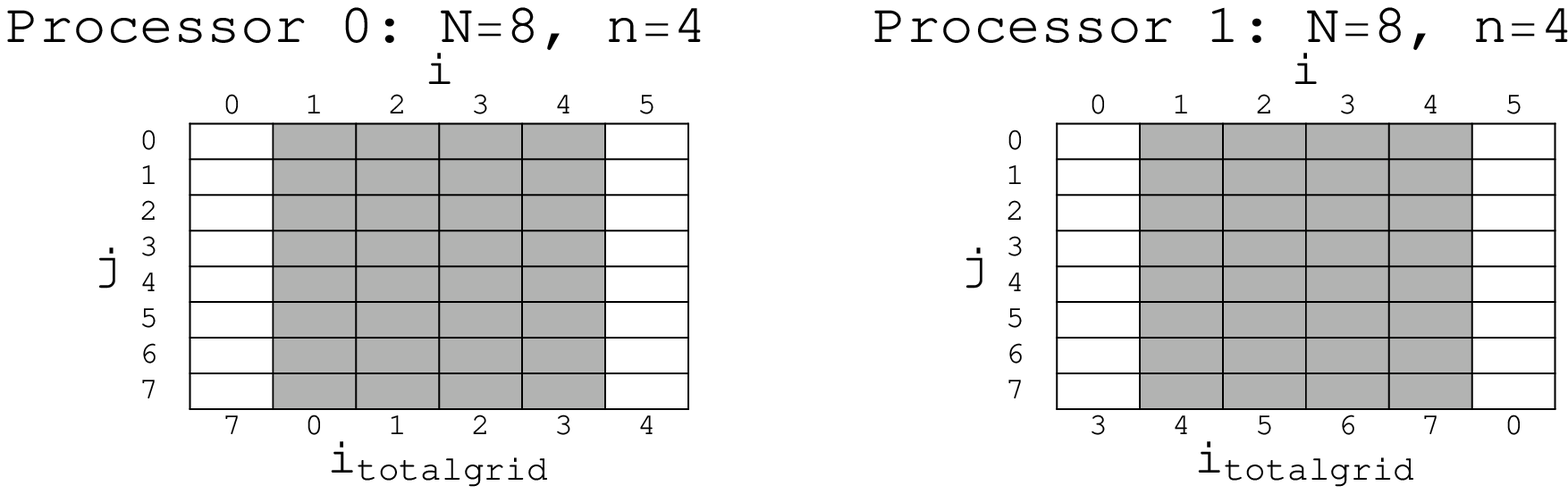}
\caption{Data layout in CLUSTEREASY}\label{clusterimage}
\end{figure}

This scheme is shown in Figure~\ref{clusterimage}. At each time step
each processor advances the field values in the shaded region, using
the buffers to calculate spatial derivatives. Then the processors
exchange edge data. At the bottom of the figure I've labeled the $i$
value of each column in the overall grid. During the exchange
processor 0 would send the new values at $i_{totalgrid}=0$ and
$i_{totalgrid}=3$ to processor 1, which would send the values at
$i_{totalgrid}=4$ and $i_{totalgrid}=7$ to processor 0.

The actual arrays allocated by the program are even larger than this,
however, because of the extra storage required by the Fourier
Transform routines. In two and three dimensions CLUSTEREASY uses
FFTW. When you Fourier Transform the fields the Nyquist modes are
stored in extra positions in the last dimension, so the last dimension
is $N+2$ instead of $N$. The total size per field of the array at each
processor is thus typically $n+2$ in 1D, $(n+2) \times (N+2)$ in 2D
and $(n+2) \times N \times (N+2)$ in 3D. In 2D FFTW sometimes requires
extra storage for intermediate calculations as well, in which case the
array may be somewhat larger than this, but usually not much. This
does not occur in 3D.

\section{Conclusions}

We have found that the speed of the simulation scales roughly as the
number of processors, provided that number is significantly smaller
than $N$, the number of gridpoints along each edge of the lattice. A
good rule of thumb is that you probably won't get much benefit from
using more processors than $N/4$. Also, you will get slightly better
performance per processor if the number of processors is a factor of
$N$ so that the processors can divide the lattice up evenly.

CLUSTEREASY offers the opportunity to do simulations of much larger,
more complex, and more realistic early universe theories than was
possible with serial simulations. We offer it in the hope that it will
be useful to the research community.

\section*{Acknowledgements}

The usual thanks offered in a paper are inadequate for the impressive
contributions made by students and other collaborators on this
project. Significant work on the project was done in particular by
Sirein Awadalla, Douglas Swanson, and Hal Finkel. Others who
contributed include Jing Li, Dessislava Michaylova, and Olga Navros. I
would also like to thank Richard Easther for valuable discussions and
the Canadian Institute for Theoretical Astrophysics for hosting me
during part of this work. Finally, I would like to thank Igor Tkachev,
my collaborator on LATTICEEASY, without whom none of this would have
been possible. G.F. was supported by NSF grant PHY-0456631.

\end{document}